\documentclass[amsmath,amssymb,pra,showpacs]{revtex4}
\pdfoutput=1
\usepackage{setspace}
\usepackage{graphicx}
\usepackage{amsmath}
\usepackage{amssymb}
\usepackage{latexsym}%
\usepackage{makeidx}         

\begin{document}

\title{On the problem of van der Waals forces in dielectric media}

\author{Lev P.~Pitaevskii}

\affiliation{
INO-CNR BEC Center and Dipartimento di Fisica, Universit\'{a} ~di Trento, 38123
Povo, Trento, Italy; \\
Kapitza Institute for Physical Problems, ul. Kosygina 2, 119334 Moscow,
Russia. \\}


\begin{abstract}
A short review of the problems which arise in the generalization of the Lifshitz theory of  van der Waals force in the case of forces inside dielectric media is presented, together with some historical remarks. General properties of the stress tensor of equilibrium electromagnetic field in media are discussed, and the importance of the conditions of mechanical equilibrium is stressed. The physical meaning of the repulsive van der Waals interaction between bodies immersed in a liquid is discussed.
\end{abstract}
\pacs{34.35.+a,42.50.Nn,12.20.-m}
\maketitle

\section{Introduction}

The quantum theory of the long range van der Waals interaction between neutral objects has a long and instructive history \cite{footnote1}. The existence of  \textit{attractive} forces between atoms was established by van der Waals on the basis of analysis of experimental data on equations of state of real gases. 

It was F.~London who understood the electric nature of these forces and built in 1930 a seminal quantum-mechanical theory of the forces at large distances, deriving the famous $1/R^6$ law for the energy of interaction \cite{London30}.  

The next important step was performed by Casimir and Polder. Using quantum electrodynamics, they derived a more general expression for the energy of the atom-atom interaction and showed that, due to retardation effects, London's law at large distances is changed to a $1/R^7$ law\cite{CP48}. In the same paper the  authors  considered for the first time the problem of including a macroscopic body. They calculated a force between an atom and perfect metal plate. The interaction between perfect metal plates was calculated by Casimir \cite{Casimir48}. 

The most general theory of the van der Waals interaction between any condensed bodies was developed by E.~Lifshitz in 1954-1955 \cite{Lifshitz54,Lifshitz56}. His theory is applicable to a body with arbitrary dielectric properties at any temperature. It also automatically takes into account relativistic retardation effects. To calculate forces one must know the dielectric and magnetic permeabilities of the bodies and  solve the Maxwell equations for their given configuration. 

It was assumed in the Lifshitz theory (as well as in previous theories) that the space between the bodies was a vacuum. A generalization of the theory where the gap between bodies is filled with some medium was a natural next step. However, to make this step one must overcome
some essential difficulties. To clarify the nature of these difficulties, let us recall the main points of the Lifshitz approach. 
This approach is based on averaging of the Maxwell stress tensor in vacuum, 
\begin{equation}
\sigma_{ik}^{(vac)}=
\frac{
E_{i}E_{k}+B_{i}B_{k}\; ,
}{4\pi }-
\frac{
E^{2}+B^{2}
}{8\pi }\delta _{ik}.  \label{vac}
\end{equation}
with respect to electromagnetic fluctuations in thermodynamic equilibrium. 

Because
the theory of equilibrium electromagnetic fluctuations in arbitrary media was already developed by Rytov \cite{Rytov} and Landau and Lifshitz \cite{LL8}, the tensor could be averaged for equilibrium conditions and the forces calculated.

The Rytov theory has a semi-phenomenological character. Rytov considered the fluctuations as created by the Langevin-like sources, namely fluctuating electric and magnetic polarizations. It was assumed that these polarizations at two points
$\mathbf{r_1}$ and   $\mathbf{r_2}$ of a medium  are correlated only when the two points coincide, i. e. their correlation functions are proportional to $\delta(\mathbf{r_1}-\mathbf{r_2})$. The coefficients of proportionality were chosen to obtain the correct density of black-body radiation from the bodies. During preparation of the book \cite{LL8}, Landau and Lifshitz derived equations of the Rytov theory using the exact fluctuation-dissipation theorem, established by Callen and Welton in 1951.

It was natural to think that the generalization for the case of bodies separated by a medium could be obtained
if one could find a general expression for the electromagnetic stress tensor for arbitrary time-dependent fields in a medium. Because the Rytov theory
describes electromagnetic fluctuations also inside a medium, it would then be possible to perform the average of the tensor.  
Thus the first step was to calculate the tensor. Note that this problem was formulated in the book \cite{LL8}, and Landau believed that it could be solved. One can read in the end of \S 61:

``Considerable interest attaches to the determination of the (time) average stress tensor giving the forces on matter in a variable electromagnetic field. The problem is meaningful for both absorbing and non absorbing media, whereas that concerning the internal energy
can be proposed only if absorption is neglected. The corresponding formulae, however, have not yet been derived.''

I had an opportunity to  read proofs of the book when I joined the 
Landau department in the Institute for Physical Problems in Moscow as a Ph.D. student in 1955. 
After reading this paragraph I decided to derive ``the corresponding formulae'' for the tensor. It was, of course, a quite ambitious  goal. After approximately  three months of  work I
met Igor Dzyaloshinskii. Because the authors asked both him and me to help in the proof-reading of Ref. \cite{LL8}, we naturally discussed   the topics of the book and I discovered that Dzyaloshinskii had also been working on the tensor problem even longer than I. We decided to join our efforts.

Our attempts were based on the use of  the second-order quantum mechanical perturbation theory to calculate the quadratic contribution of the fields into the tensor. 
One can obtain formal equations; however, for a medium with dissipation we could not express  them in terms of dielectric and magnetic permeabilities. Oddly enough, one could easily obtain an equation for the \textit{total} force acting on a body in vacuum. However, trying to obtain from this equation the \textit{force density} inside the body, one necessary violated the condition of the symmetry of the stress tensor, or other conditions implied for  the tensor. I also tried to develop a thermodynamic approach. However,  the entropy increase due to the dissipation makes this approach meaningless. 

We worked long enough and finally decided that the problem was hopeless. (In any case it was my opinion.)

However, after a period of disappointment, I suddenly recognized that the problem of the van der Waals forces for the case of a liquid film   can be solved without involving the general tensor problem. Because, as we will see below, these considerations actually were not employed and never were published, I would like to present them here. 

Let us consider a  liquid film of thickness $d$ in vacuum. Its free energy per unit of area can be presented in the form
\begin{equation}
{\cal F}(T,d)=\varphi_0(T)d+{\cal F}^{(elm)}(T,d)  \label{F}
\end{equation}
where $\varphi_0(T)$ is the density of free energy for a bulk body and ${\cal F}^{(elm)}(T,d)$ is the contribution of the van der Waals forces, which can be normalized in a such way that ${\cal F}^{(elm)} \to 0$ at
$d \to \infty$. The goal of the theory in this case is to calculate ${\cal F}^{(elm)}(T,d)$. However, this quantity can be calculated on the basis of the Lifshitz theory. Let us consider our film at the distance $l$ from a half-space  \textit{of the same liquid} (see Fig.~1).
\begin{figure}[t]
\includegraphics[width=7 cm]
{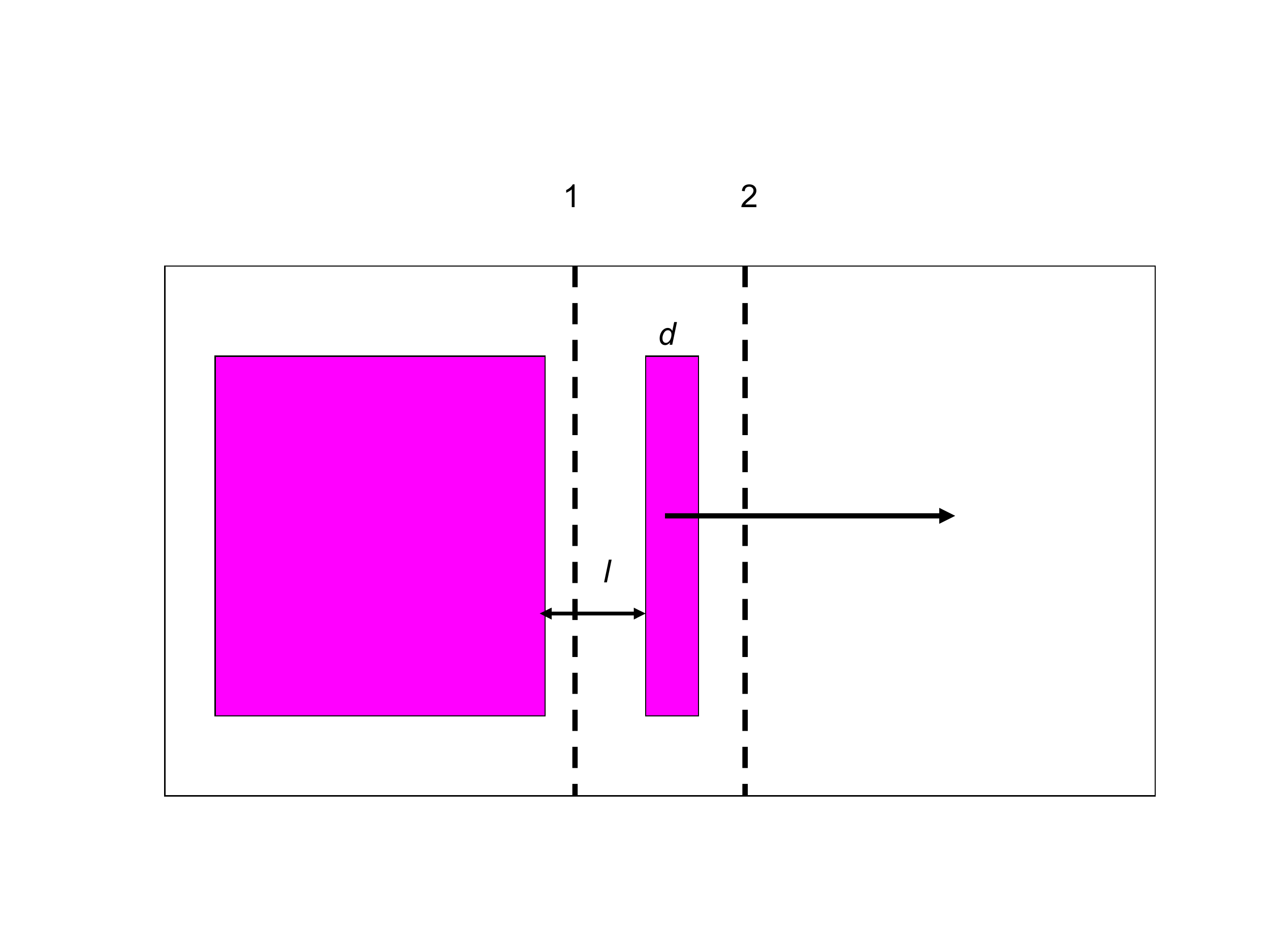}
\caption{ Scheme for calculation of the chemical potential of a film in terms of the stress tensor in vacuum.
}
\label{fig:film}
\end{figure}
The theory permits the calculation of the force $F(d,l)$ between the film and the half-space and the change of the free energy when the film moves from the surface of the bulk liquid to infinity:
\begin{equation}
\Delta {\cal F}(T,d)=\int_0^{\infty}Fdl \label{DF}\; .
\end{equation}
This quantity is just  ${\cal F}^{(elm)}(T,d)$. Indeed, if the film is near the surface of the half space, the configuration corresponds to a bulk body. Actually, the integral (\ref{DF}) is divergent at small $l$. However, this infinite contribution does not depend on $d$ and can be omitted altogether with a $d$-independent constant.

We discussed the  idea with Landau and he agreed with the argumentation. We immediately began calculations of the force $F$ for the three-boundary configuration of Fig.~1. E. M. Lifshitz, who was interested in the film problem very much, joined us. However, Landau cooled our ardor. He said that it is meaningless to solve a three boundary problem to find an answer for a two-boundary one and that the integration (\ref{DF}) must be performed in a general form, ``in  some symbolic way'' as he said.  

As a result, Dzyaloshinskii and I began to look for a different approach, being sure this time that the answer exists. It was a lucky coincidence that just in this period Dzyaloshinskii, in collaboration with Abrikosov and Gorkov, worked on developing the Matsubara diagram technique for the solution of \textit{equilibrium} problems in the quantum many-body theory. We decided to try this approach. The first attempt was very successful. We immediately recognized  which diagrams are important. As an intermediate  result we obtained an equation where the van der Waals contribution to the free energy was expressed in terms of an integration with respect to the charge of the electron. This equation was correct, but not very useful because the dielectric permeability can, of course, be measured only for the actual value of the charge. Finally, however, we discovered that in the Matsubara technique \textit{the variation} of the free energy with respect to the dielectric permeability can be expressed in terms of the Matsubara Green's function of imaginary frequency. This permitted us to calculate the tensor explicitly \cite{DP}.

The reason why our previous attempts were in vain became clear now. One can obtain the tensor of the \textit{equilibrium} electromagnetic field in a medium, that is, the tensor of van der Waals forces, but not the tensor of arbitrary electromagnetic fields. 
The possibility to derive the tensor of the equilibrium fluctuation electromagnetic field in an absorbing medium does not mean, of course, the possibility to determine the tensor for an arbitrary variable field. Even if this quantity has physical meaning, according to the Landau conjecture quoted above, it does not mean that it can be expressed in terms of the electric and magnetic permeabilities $\varepsilon(\omega)$ and $\mu(\omega)$. I believe that this is impossible. In any case, the tensor, obtained by direct calculation for a plasma, where the problem can be  explicitly solved using the Boltzmann kinetic equation, cannot be expressed in such a way \cite{Perel68}. The difficulty, of course, is the energy dissipation. In a transparent medium, where dissipation is absent, the tensor can be obtained for arbitrary  non-equilibrium fields\cite{Pitaevskii61}. 

Before discussing concrete results, let us discuss general properties of the stress tensor in thermal equilibrium. It can be presented in the form
 \begin{equation}
\sigma_{ik}=-P_{0}(T,\rho)\delta _{ik}+
\sigma_{ik}^{(elm)} \; ,
 \label{gen}
\end{equation}
where $P_{0}(T,\rho)$ can be defined as the pressure of a uniform infinite liquid at given density $\rho$ and temperature $T$  and  
$\sigma_{ik}^{(elm)}$  is the contribution from the electromagnetic fluctuations, i.e., the van der Waals interaction. This contribution must satisfy several important conditions:
 
 i) The tensor must be symmetric: $\sigma_{ik}^{(elm)}=\sigma_{ki}^{(elm)}$.
 
 ii)  The van der Waals part of the force, acting on the liquid, must be derivable from a potential:
 \begin{equation}
F^{(elm)}_{i}=\partial_{k}
\sigma_{ik}^{(elm)} =-\rho \partial_i \zeta^{(elm)} \; .
 \label{Fgen}
\end{equation}
Actually $\zeta^{(elm)} $ is just the contribution of the van der Waals interaction to the chemical potential of the fluid. 
The first condition is a direct consequence of the symmetry of the microscopic energy-momentum tensor. The tensor $(-\sigma_{ik})$ is its averaged spatial part. Condition ii)  ensures the possibility of mechanical equilibrium of the fluid in the presence of the van der Waals interaction.  Indeed, the condition for such an equilibrium is $F_{i}=\partial_{k} \sigma_{ik}=0$. Taking into account that $dP_0=\rho d\zeta_0(\rho,t)$, we can rewrite this equation as 
\begin{equation}
\partial_{i} \zeta_0- F^{(elm)}_{i}/\rho=0 \;. 
\label{eq0}
\end{equation}
which implies  (\ref{gen}) for arbitrary configurations of interactive bodies.
Thus violation of the condition ii)  would result in permanent flow of the liquid in equilibrium and actually would permit us to build the notorious \textit{Perpetual Motion } machine.

For an analogous reason,  
on the boundary between a fluid and a solid the tangential components of the tensor must be continuous. This condition is satisfied automatically by virtue of the boundary conditions for the fluctuating fields. If the normal to the surface is directed along $z$, it must be the case
that $\sigma_{\alpha z}^{elm(1)}=\sigma_{\alpha z}^{elm(2)}, \alpha=x,y.$
Violation of this condition would result in the existence of permanent flow of the liquid near a solid boundary.

Equation (\ref{eq0}) can be written as a condition of constancy of the total chemical potential $\zeta$:
\begin{equation}
\zeta(\rho,T)=\zeta_0(\rho,t)+ \zeta^{(elm)}(\rho,T)=const \;.
\label{equzeta}
\end{equation}
Notice that,  neglecting the change in density of the liquid under the influence of the van der Waals forces, one can express
this condition also as
\begin{equation}
P_{0}(\rho,T)/\rho+ \zeta^{(elm)}=const \;.
\label{equ1}
\end{equation}
For calculations of the forces in the state of mechanical equilibrium one can omit this part of the tensor and exclude the pressure, i. e. use instead of (\ref{gen}) the tensor
 \begin{equation}
{\sigma}\prime_{ik}= 
\sigma_{ik}^{(elm)}+ \rho \zeta^{(elm)}(T,\rho)\delta_{ik} \; . 
 \label{equ}
\end{equation} 
Notice that the tensor ${\sigma}\prime_{ik}$ \textit{by definition} satisfies the equation 
 \begin{equation}
\partial_{k}{\sigma}\prime_{ik}=0 \; .
\label{ds0}
\end{equation} 
This obvious property results sometimes in misunderstandings (Ref. \cite{Comment}).

  
\section{Free energy of the equilibrium electromagnetic field in an absorbing medium}

I present now a simplified version of our deviation of the force tensor in a liquid obtained for the first time in \cite{DP}. It is sufficient to take into account only the electromagnetic interaction in the system. Nuclear forces are obviously irrelevant to our problem. Then the correction due to interaction to the free energy can be presented in the Matsubara technique as a set of  ``ring'' diagrams (see Fig. 2), where the dashed lines represent the Matsubara Green's functions of the electromagnetic field ${\cal D}_0$ without interaction\cite{AGD}. Every ``bead'' is the polarization operator $\Pi$, which includes all diagrams which cannot be separated into parts, connected by one dashed line. It is important that the diagrams of Fig. 2 cannot be summed up into the exact Green's function, because of the extra factor $1/n$ in each term, where $n$ is the number of the ''beads''.

\begin{figure}[t]
\includegraphics[width=10 cm]
{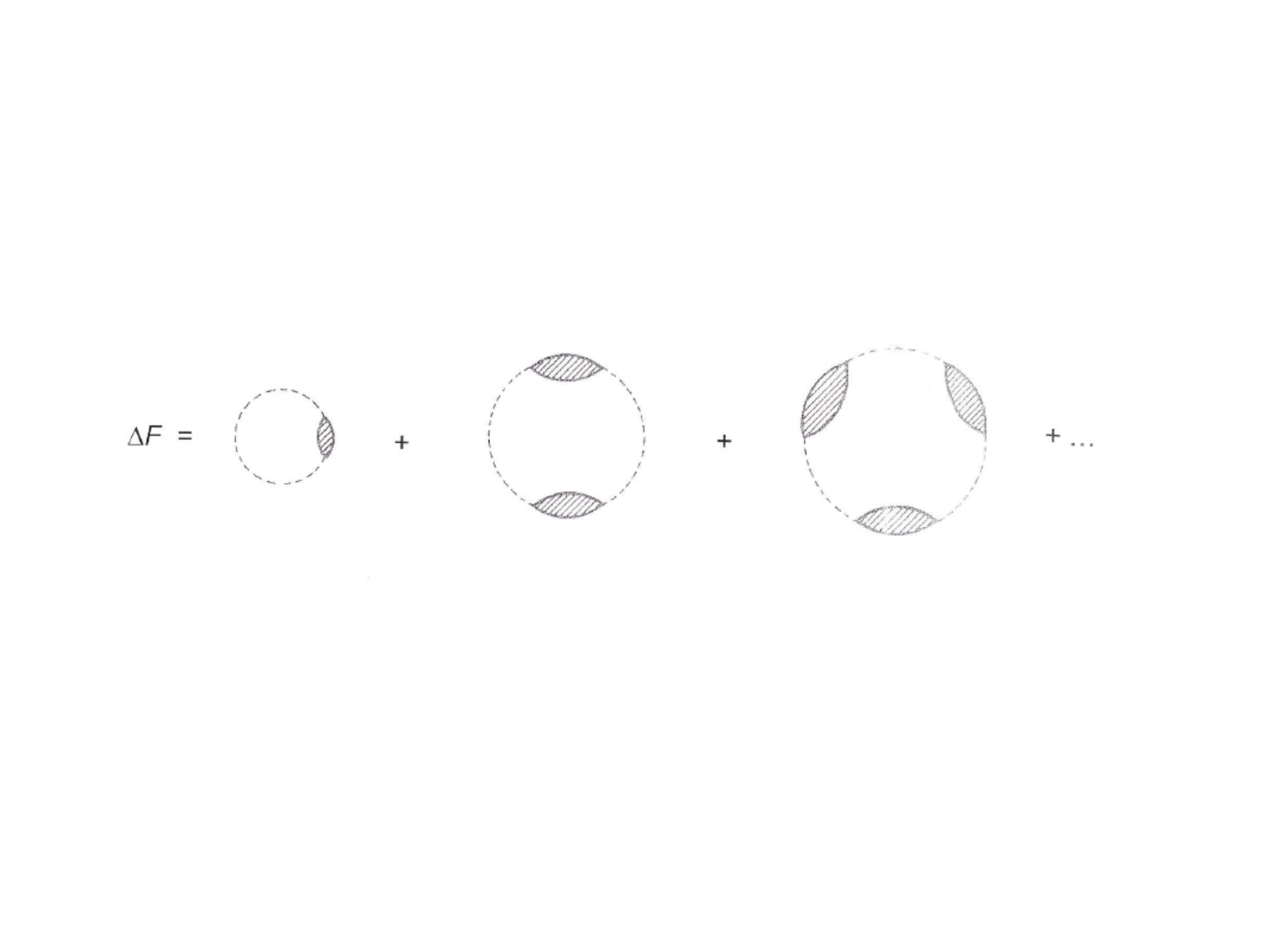}
\caption{ Diagrammatic representation of  corrections to the free energy.
}
\label{fig:diag}
\end{figure}
 
Let us calculate now the variation of the free energy with respect to a small  change $\delta \Pi$ of the density of the liquid. The crucial point is that the factor $1/n$ will be canceled as a result of the variation and the parts of the diagrams with the non-variated beads will be summed up to the exact Green's function. The result is
 \begin{eqnarray}
\delta {\cal F} = \delta {\cal F}_0 - {\sum_{s=0}^{\infty}}~^\prime \int{\cal D}_{ik}(\mathbf{r},\mathbf{r}';\xi_{s})\delta \Pi_{ik}(\mathbf{r},\mathbf{r}';\xi_{s})d\mathbf{r}d\mathbf{r}'\; ,
 \label{deltaF}
\end{eqnarray}
where $\xi_s=2s\pi T$ and the term with $s=0$ is taken with a factor (1/2)\cite{footnote}. This is  \textit{an exact} equation of quantum electrodynamics. It is valid also in a vacuum, where $\Pi$ describes the radiation correction to ${\cal D}$. However, it is practically  worthless, because explicit expressions for ${\cal D}$ and $\Pi$ cannot be obtained. It also contains ultraviolet divergences. However, these divergences are due to contributions from the short wave-length fluctuations, whereas we are interested in effects due to the inhomogeneity of the medium. i. e., the presence of boundaries of bodies etc. This permits us to produce a renormalization of this equation. To do this, let us write the ${\cal D}$-function as 
\begin{eqnarray}
{\cal D}_{ik}(\mathbf{r},\mathbf{r}';\xi_{s}) = \left [ {\cal D}_{ik}(\mathbf{r},\mathbf{r}';\xi_{s}) - \bar{{\cal D}}_{ik}(\mathbf{r},\mathbf{r}';\xi_{s}) \right ]+\bar{{\cal D}}_{ik}(\mathbf{r},\mathbf{r}';\xi_{s}) \; ,
 \label{ren}
\end{eqnarray}
where $\bar{{\cal D}}_{ik}(\mathbf{r},\mathbf{r}';\xi_{s})$ is the Green's function of an auxiliary homogeneous infinite medium whose permeabilities are the same as that of the actual medium at the point $\mathbf{r}'$. After  substitution into (\ref{deltaF}) the third term can be absorbed in the term $\delta {\cal F}_{0}$.  This term acquires the meaning of the variation of the free energy of this uniform medium. 
Then one gets instead (\ref{deltaF}):
\begin{eqnarray}
\delta {\cal F} = \delta {\cal F}_0 - {\sum_{s=0}^{\infty}}~^{\prime} \int\left [ {\cal D}_{ik}(\mathbf{r},\mathbf{r}';\xi_{s}) - \bar{{\cal D}}_{ik}(\mathbf{r},\mathbf{r}';\xi_{s}) \right ]\delta \Pi_{ik}(\mathbf{r},\mathbf{r}';\xi_{s})d\mathbf{r}d\mathbf{r}'\; .
 \label{dF1}
\end{eqnarray}
In equation (\ref{dF1}) the important fluctuations are those whose wavelengths are of the same order of magnitude as the inhomogeneities of the system (e.g.,
the thickness of  films and separations of  bodies). These lengths are assumed to be large compared to interatomic dimensions. However, these long-wavelength fluctuations can be described by macroscopic electrodynamics. According to the general theory, the Matsubara Green's function
${\cal D}_{ik}(\mathbf{r},\mathbf{r}';\xi_{s})=D_{ik}^{R}(\mathbf{r},\mathbf{r}';i\xi_{s})$, where $D^{R}$ is the ``usual'' retarded Green's
function for the vector-potential of the electromagnetic field. Accordingly, ${\cal D}$ satisfies in the macroscopic limit the explicit equation
 \begin{eqnarray}
\left [\partial_i\partial_l -\delta_{il}\Delta+ (\xi_{s}^2/c_2)\varepsilon(i|\xi_s|,\bf{r})\delta_{il}\right ]{\cal D}_{ik}(\mathbf{r},\mathbf{r}';\xi_{s})=-4\pi\hbar\delta_{ik}\delta(\bf{r}-\bf{r}')\; .
 \label{D}
\end{eqnarray}  
The equation for $\bar{\cal D}$ can be obtained by changing in (\ref{D}) the permeability $\varepsilon(i|\xi_{s}|,\bf{r})$ to $\varepsilon(i|\xi_{s}|,\bf{r}')$. However, in the majority of practical problems the excluding of the divergences can be achieved simply by omitting terms which do not depend of the spatial parameters, e.g., on the distances between bodies. Taking this into account, and to avoid complications of the equations, we will denote in the future the difference (\ref{ren}) as ${\cal D}$. It is worth noting that the left-hand side of (\ref{D}) coincides with the Maxwell equations for the vector potential $\bf{A}$ of the electromagnetic field with the frequency $i\xi_{s}$ in the gauge where the scalar potential $\phi=0$. The final results of the theory do not depend, of course, on the gauge. The Green's function ${\cal D}_0$ satisfies the same equation (\ref{D}) with $\varepsilon =1$. Let us write ({\ref{D}}) symbolically as
$\hat {\cal D}^{-1}{\cal D} =\delta_{il}\delta(\bf{r}-\bf{r}')$ and the equation 
for  ${\cal D}_0$ as $\hat {\cal D}_0^{-1}{\cal D}_0 =\delta_{il}\delta(\bf{r}-\bf{r}')$. Then from the definition
$\Pi_{ik}=\left [ {\cal D}^{-1}_{0ik}
-{\cal D}^{-1}_{ik} \right ]$ we find an equation for $\Pi$:
\begin{eqnarray}
 \label{Pi}
\Pi_{kl}(\xi_{s};\bf{r}_1,\bf{r}_2)=\frac{\xi_{s}^2}{4\pi}\delta_{kl}\delta(\bf{r}-\bf{r}')\left [\varepsilon(i\xi_{s};\bf{r}_1)-1 \right ] \;.
\end{eqnarray}
This permits us to express the variation of the free energy in  terms of $\varepsilon$:
\begin{gather}
\delta {\cal F}= \delta {\cal F}_0 - \frac{T}{4\pi} {\sum_{s=0}^{\infty}}~^{\prime} \int\left [\xi^2_s {\cal D}_{ll}(\mathbf{r},\mathbf{r};\xi_{s}) 
\right 
]\delta \varepsilon(i\xi_{s};\mathbf{r})d\mathbf{r}\; .
 \label{dFfin}
\end{gather}
It is convenient for further equations to introduce a new function
\begin{gather}
 {\cal D}^E_{ik}(\mathbf{r},\mathbf{r}';\xi_{s}) =-\xi^2_s
{\cal D}_{ik}(\mathbf{r},\mathbf{r}';\xi_{s}) \; .
 \label{DE}
\end{gather}
Function ${\cal D}$ describes  the quadratic fluctuations of the vector potential in the Matsubara
technique. Accordingly, the function ${\cal D}^E$ describes fluctuations of the electric field. We will also use the function 
\begin{gather}
 {\cal D}^H_{ik}(\mathbf{r},\mathbf{r}';\xi_{s}) ={\rm curl}_{il}{\rm curl}'_{km}
{\cal D}_{lm}(\mathbf{r},\mathbf{r}';\xi_{s}) 
 \label{DE}
\end{gather}
describing fluctuations of the magnetic field.
Now we can rewrite (\ref{DE}) as
\begin{gather}
\delta {\cal F} = \delta {\cal F}_0 + \frac{T}{4\pi} {\sum_{s=0}^{\infty}}~^{\prime} \int{\cal D}^{E}_{ll}(\mathbf{r},\mathbf{r};\xi_{s}) 
\delta \varepsilon(i\xi_{s};\mathbf{r})d\mathbf{r}\; .
 \label{dFfin1}
\end{gather}


\section{Stress tensor of the van der Waals interaction inside an absorbing medium.}

We can use equation (\ref{dFfin1}) to calculate the tensor of van der Waals forces in a fluid. It is instructive, however, as a first step to compare the equation for a free energy variation for given sources of field in a dielectric in the absence of dispersion (Ref. \cite{LL8}, equation (15.19)):
\begin{gather}
\delta {\cal F} = \delta {\cal F}_0 - \int\frac{E^2}{8\pi}\delta\varepsilon d\mathbf{r}
 \label{dFstat}
\end{gather}
This equation permits the calculation of the force $\mathbf{f}$ and finally to find the stress tensor (see Ref. \cite{LL8}, (15.9) and (35.2), in the presence of both electric and magnetic fields we must take the sum of these equations):
\begin{gather}
\sigma_{ik}^{A}=-P_{0}\delta _{ik}+
\frac{\varepsilon E_{i}E_{k}+H_{i}H_{k}}{4\pi }%
\nonumber \\
-\frac{E^{2}}{8\pi }\left[ \varepsilon -\rho \left( \frac{\partial
\varepsilon }{\partial \rho }\right) _{T}\right] \delta _{ik}-\frac{H^{2}}{%
8\pi }\delta _{ik}.  \label{Abr}
\end{gather}
This equation was derived by M. Abraham around 1909 and is
one of the most important results of the electrodynamics of continuous media.

Now we can write the tensor of the van der Waals forces by direct analogy with (\ref{Abr}). Indeed, the functions ${\cal D}^E_{ik}(\mathbf{r},\mathbf{r}';\xi_{s})$ and ${\cal D}^H_{ik}(\mathbf{r},\mathbf{r}';\xi_{s})$ satisfy equations which are similar to the products $E_{i}(\mathbf{r})E_k(\mathbf{r}')$ and $H_{i}(\mathbf{r})H_k(\mathbf{r}')$. The presence of the $\delta$-function term on the right-hand side of (\ref{D}) is not important because this term in any case will be eliminated in the course of the renormalization.
Thus, the general expression for the stress
tensor for a fluid with $\mu =1$  is \cite{DP}:
\begin{gather}
\sigma_{ik}=-P_{0}\delta _{ik}-\frac{\hbar T}{2\pi }\Big \{
 \sum_{s=0}^{\infty}~^{\prime}\Big(
\varepsilon D_{ik}^{E}+D_{ik}^{H}\nonumber \\
-\frac{1}{2}D_{ii}^{E}\Big [ \varepsilon
-\rho \Big ( \frac{\partial \varepsilon }{\partial \rho }\Big ) _{T}\Big ]
\delta _{ik}-\frac{1}{2}D_{ii}^{H}\delta _{ik}\Big ) \Big \} ,  \label{DP}
\end{gather}
where $D_{ik}^{E}$ and $D_{ik}^{H}$  were defined above,
 $\varepsilon =\varepsilon \left(
\rho ,T,i\zeta _{s}\right) ,\zeta _{s}=2sT/\hbar $, and $P_{0}\left( \rho
,T\right) $ is the pressure as a function of density and temperature in the
absence of an electric field. 

Equation $\left( \ref{DP}\right) $ assumes the
system to be in thermal, but still not in {\em mechanical } equilibrium. As  was said before, the last condition can be formulated as a condition of the constancy of the chemical potential $\zeta$, which can be defined by the equation $\delta F=\int \zeta \delta  \rho d\mathbf{r}$. The variation must be taken at fixed boundaries of the bodies. One has from (\ref{dFfin1})
\begin{gather}
\zeta(\rho, T) = \zeta_0(\rho,T) + \frac{\hbar T}{4\pi} {\sum_{s=0}^{\infty}}~^{\prime} {\cal D}^{E}_{ll}(\mathbf{r},\mathbf{r};\xi_{s}) 
\frac{\partial \varepsilon(i\xi_{s};\mathbf{r})}{\partial \rho}\; .
 \label{zeta}
\end{gather}

The condition  of   mechanical equilibrium  means that 
$\zeta(\rho, T)=const$. Let the fluid have  uniform density  in the
absence of the van der Waals forces. Taking into account that $dP_0=\rho d\zeta_0$ and neglecting in the second term
any change of $\rho$ due the van der Waals interaction, we can rewrite 
the condition of equilibrium as: 
\begin{equation}
P_{0}+\frac{\hbar T}{4\pi }\sum\limits_{s}~^{\prime}D_{ii}^{E}\rho \left( \frac{%
\partial \varepsilon }{\partial \rho }\right) _{T}=const \; .  \label{mech}
\end{equation}
This equation can be used to calculate the perturbation $\delta \rho$ of the density of the liquid due to the
van der Waals forces. Expanding the first term with respect to $\delta \rho$, we easily find
\begin{equation}
\delta \rho=-\frac{\hbar T}{4\pi }\sum \limits_{s}~^{\prime}D_{ii}^{E}\rho \left( \frac{%
\partial \varepsilon }{\partial P }\right) _{T} \; .  \label{drho}
\end{equation}
Equation (\ref{mech})
implies that a part of the stress tensor $\left( \ref{DP}%
\right) $ is constant through the fluid, being a uniform compressing or
expanding pressure. This part can be omitted in many problems, for example
in the calculation of the full force acting on a body embedded in the fluid.
Subtracting the constant tensor $\left[ -P_{0}-\frac{\hbar T}{4\pi }%
\sum\limits_{n}D_{ii}^{E}\rho \left( \frac{\partial \varepsilon }{\partial
\rho }\right) _{T}\right] \delta _{ik}$ from $\left( \ref{DP}\right) $, one
arrives to the  ``contracted'' tensor, which was obtained for the first time  in \cite{DLP} (see also Ref. \cite{DLP2}): 
\begin{equation}
\sigma_{ik}^{\prime}=-\frac{\hbar T}{2\pi }\left\{ \sum_{s}~^{\prime}\left( \varepsilon
{\cal D}_{ik}^{E}+{\cal D}_{ik}^{H}-\frac{1}{2}\varepsilon {\cal D}_{ii}^{E}\delta _{ik}-\frac{1}{%
2}{\cal D}_{ii}^{E}\delta _{ik}\right) \right\}  \; . \label{elmM}
\end{equation}

I would like to stress that the ``$P_0$'' term in the tensor (\ref{DP}) plays an important role. Ignoring this term
would lead to wrong results. In this connection it is appropriate  to quote Landau and Lifshitz's remark (see Ref. \cite{LL8}, \S 15):
``The problem of calculating the forces (called \textit{pondermotive} 
forces) which act on a dielectric in an arbitrary non-uniform electric field is fairly complicated...''

Notice that $\partial \sigma_{ik}^{\prime}/\partial x_{k}=0$ and hence $\oint
\sigma_{ik}'dS_{k}=0$ for integration over any closed surface, surrounding a
volume of a uniform fluid, just due to the fact that in mechanical
equilibrium electromagnetic forces are compensated by a pressure gradient.
Analogous integration over any surface surrounding a solid body  gives the total force acting on the body.

Equivalent theories of the force between bodies separated by a liquid were developed
by Barash and Ginzburg \cite{BG} and Schwinger, DeRead and Milton \cite{SDM}.   The method of  \cite{BG} is based on a very interesting and new physical idea. I cannot discuss it here. Notice only, that the method permits us to calculate forces on the basis of the solution of the imaginary
frequencies ``dispersion relation'' ${\cal D}^{-1}(i\xi_s)=0$, without an actual calculation of ${\cal D}$. This results in further simplification of calculations. The authors of \cite{SDM} performed the free energy variation assuming actually the condition of the mechanical equilibrium from the very beginning and obtained directly equation (\ref{elmM}).


\section{Van der Waals forces between bodies separated by a liquid}

Now we can calculate the force acting on bodies separated by a dielectric liquid. It is worth  noticing, however, that even for bodies in vacuum the method, based on using the imaginary-frequencies Green's functions, involves simpler calculations than the original Lifshitz method, because the solution of the equation for the Green's functions is simpler than the procedure of averaging  of the stress tensor. 

It was shown in \cite{DLP} that if the problem has been solved for bodies in a vacuum, the answer for bodies in liquid can be found by a simple scaling transformation. Let us denote the dielectric permeability of the liquid as $\varepsilon$. If we perform a coordinate transformation $\mathbf{r}=\tilde{\mathbf{r}}/{\varepsilon}^{1/2}$ and introduce the new functions ${\cal D}_{ik}=\tilde{{\cal D}}_{ik}/\varepsilon^{1/2}$ and ${\cal D}^E_{ik}=\tilde{{\cal D}}^E_{ik}\varepsilon^{1/2}$,
 ${\cal D}^H_{ik}=\tilde{{\cal D}}^H_{ik}\varepsilon^{3/2}$, then
 \begin{equation}
\sigma_{ik}^{\prime}=-\frac{\hbar T}{2\pi }\left\{ \sum_{s=0}^{\infty}~^{\prime}{\varepsilon}^{3/2}\left( 
\tilde{{\cal D}}_{ik}^{E}+\tilde{\cal D}_{ik}^{H}-\frac{1}{2} \tilde{\cal D}_{ii}^{E}\delta _{ik}-\frac{1}{%
2}\tilde{\cal D}_{ii}^{E}\delta _{ik}\right) \right\}  \; . \label{stilde}
\end{equation}
One can see easily that the new functions $\tilde{{\cal D}}_{ik}$ satisfy in the new coordinates $\tilde{\mathbf{r}}$ equations of the same form
(\ref{D}) for bodies in vacuum, while the permeabilities $\varepsilon_{\alpha}$ of the bodies were changed to $\varepsilon_{\alpha}/\varepsilon$. 

One can usually neglect the influence of the temperature on the forces between bodies in a liquid. Then one can change 
$T{\sum}_{s=0}^{\infty}~^{\prime}... \to
\frac{\hbar}{2\pi}\int_{0}^{\infty}....d\xi_s$. We will consider below only this case. We also will consider only the small ``London'' distances
where the characteristic distance between bodies $l \ll \lambda$, where $\lambda $ is the characteristic wavelength of the absorption spectra of the media. In this case one can neglect the magnetic Green's function ${\cal D}_{ik}^{H}$ and the electric function can be presented as
${\cal D}_{ik}^{E}=\hbar\partial_i\partial_k\prime\phi$, where the ``electrostatic'' Green's function $\phi$
satisfies equation \cite{Volokitin,Pitaevskii08}
\begin{equation}
\partial_i[\varepsilon(i\xi;\mathbf{r})\partial_i\phi(\xi;\mathbf{r},\mathbf{r}')] =-4\pi\delta(\mathbf{r}-\mathbf{r}')
\; . \label{phi}
\end{equation}
Thus $\phi$ is just the potential of a unit charge placed at point $\mathbf{r}'$. We will present here results for two important problems.

\subsection{Interaction of a small sphere with a plane body}

As a first example we consider  a dielectric sphere in the vicinity of a plane surface of a bulk body. Let the radius $R$ of the sphere be small in comparison with the distance $l$ between the sphere and the surface.   We consider first the problem in vacuum. Then the energy of interaction can be
obtained directly from (\ref{dFfin1}), taking into account that the change of the dielectric permeability due to the presence of the sphere at point $\mathbf{r}_0$ is
$\delta \varepsilon (\omega)= 4\pi\alpha (\omega) \delta(\mathbf{r}-\mathbf{r}_0)$, where $\alpha (\omega)$ is the polarizability of the sphere.
In the zero-temperature London regime we get
\begin{equation}
V(l)=\frac{\hbar}{2\pi}\int_{0}^{\infty}\alpha(i\xi)\left [{\cal D}^E_{ll}(\xi;(\mathbf{r},\mathbf{r}')\right ]_{\mathbf{r} \to \mathbf{r} \to \mathbf{r}_0} \label{Vl}d\xi \; .
\end{equation}
The ``potential'' $\phi$ can be taken from \cite{LL8}, \S 7, Problem 1. A simple calculation then gives 
\begin{equation}
{\cal D}^E_{ll}(\xi;(\mathbf{r}_0,\mathbf{r}_0) =-\frac{\hbar}{2l^3}\frac{\epsilon_1(i\xi)-1}{\epsilon_1(i\xi)+1}\; .
\end{equation}
Taking into account that
\begin{equation}
\alpha(i\xi)
=R^3\frac{\varepsilon_2(i\xi)-1}{\varepsilon_2(i\xi)+2} \; ,\label{alpha}
\end{equation}
we find
\begin{equation}
V(l)=-\frac{\hbar R^3}{4\pi l^3} \int_{0}^{\infty}\frac{(\varepsilon_2(i\xi)-1)(\epsilon_1(i\xi)-1)}{(\varepsilon_2(i\xi)+2)(\epsilon_1(i\xi)+1)}d\xi \; .
\label{Vvac}\end{equation}
The force acting on the sphere is
\begin{equation}
F(l)=-\frac{dV}{dl}=-\frac{3\hbar R^3}{4\pi               l^4}\int_{0}^{\infty}\frac{(\varepsilon_2(i\xi)-1)(\epsilon_1(i\xi)-1)}{(\varepsilon_2(i\xi)+2)(\epsilon_1(i\xi)+1)}d\xi \; .
\label{Fvac}\end{equation}
One must be careful when rewriting this equation for a case of bodies separated by liquid. The transformation  was formulated for the tensor $\sigma'$.  Taking into account that  $F=\int \sigma'_{zz}dxdy$, we conclude that it is enough to change $\epsilon_1 \to \epsilon_1/\epsilon$ and $\epsilon_2 \to \epsilon_2/\epsilon$:
\begin{equation}
F(l)=\frac{3\hbar R^3}{4\pi l^4}\int_{0}^{\infty}\frac{(\varepsilon_2(i\xi)-\varepsilon(i\xi))(\epsilon_1(i\xi)-\varepsilon(i\xi))}{(\varepsilon_2(i\xi)+\varepsilon(i\xi))(\epsilon_1(i\xi)+\varepsilon(i\xi))}d\xi \; .
\label{Fliq}\end{equation}

\subsection{Interaction between two parallel plates}

Let us consider now the force between solid bodies 1 and 2 separated by very small distances. It should be noted that, for a rigorous statement of the problem, it is necessary to consider at least one of the bodies as being of finite size and  surrounded by the liquid. Then $F_i=\oint
\sigma_{ik}'dS_{k}$ is the total force acting on the body. However, since the van der Waals forces decrease very quickly with distance, the integrand is actually different from zero only inside the gap and the force can be calculated as  $F=F_z=\int
\sigma_{zz}'dxdy$. Notice that, due to equation (\ref{ds0}), the quantity $\sigma'_{zz}$ does not depend on $z$. Finally the force per unit area  can be expressed as\cite{DLP}
\begin{equation}
F=\frac{\hbar}{16\pi l^3}\int_0^{\infty}\int_0^{\infty}x^2\left 
[\frac{(\varepsilon_1+\varepsilon)(\varepsilon_2+\varepsilon)}{(\varepsilon_1-\varepsilon)(\varepsilon_1-\varepsilon)}e^x-1 \right ]
dxd\xi
\label{plates}
\end{equation}
where the dielectric permeabilities must be taken as functions of the imaginary frequency $i\xi$.


\section{Remarks about repulsive interactions}

It is well known that forces between bodies in vacuum are attractive. In the cases considered in the previous section it is obvious, because
for any body $\varepsilon(i\xi)>1$ for $\xi >0$. It also follows from Eqs. (\ref{Fliq}) and (\ref{plates}) that forces are attractive for  bodies of the same media ($\varepsilon_1=\varepsilon_2$).

If, however, the bodies are different, the force can be either attractive or repulsive. It is clear from (\ref{Fliq}) and (\ref{plates}) 
that if the differences $\varepsilon_1-\varepsilon$ and $\varepsilon_2-\varepsilon$ have different signs in the essential region of values $\xi$, we have $F<0$, that is, the bodies repel one another. 

To understand better the physical meaning of this repulsion, let us consider the problem of the body-sphere interaction and assume that the materials of both the sphere and the liquid (but not the body) are optically rarefied, i. e., that 
$\varepsilon_2-1 \ll 1$ and $\varepsilon -1 \ll 1$. Then (\ref{Fliq}) can be simplified as
\begin{equation}
F(l)\approx \frac{3\hbar R^3}{8\pi l^4} 
\int_{0}^{\infty}\frac{(\varepsilon_2(i\xi)-1)}{(\varepsilon_2(i\xi)+1)}(\epsilon_1(i\xi)-\varepsilon(i\xi))d\xi \; .
\label{Frar}\end{equation}
The force is now expressed as a difference of two terms, with clear physical meaning. The first term is the force that acts on the sphere in vacuum. The second term is the force that would act in vacuum  on an identical sphere, but with optical properties of the liquid. This second term is an exact analogy of the Archimedes' buoyant force, which acts on a body embedded in a liquid in a gravitational field. This remark again stresses the importance of the condition of the mechanical equilibrium in the liquid. Of course, such a simple interpretation is possible only in the limit of rarefied media. 

The existence of the repulsive van der Waals forces, predicted in \cite{DLP}  was confirmed in  several  experiments (see\cite{Capasso,Capasso2} and references therein). See also the chapter by Capasso {\it et al.} in this volume.
Corresponding experiments  are, however, quite difficult. 
Forces at large distances are quite small, while at small distances the atomic structure of the media becomes essential.


\section{Liquid films}

The van der Waals forces play an important part in the physics of surface phenomena, and in the properties of thin films in particular. 
A fundamental problem here is the dependence of the chemical potential $\zeta$ on the thickness $d$ of a film. For example, the thickness of a film on a solid surface  in equilibrium with the  vapour at pressure $P$ is given by the equation
\begin{equation}
\zeta(P,d)=\zeta_0(P)+\frac{T}{m}\ln\frac{P}{P_{sat}}
 \; .
\label{sat}\end{equation}

If the thickness $d$ of the film is large compared to interatomic distances, this dependence is defined mainly by the van der Waals forces. Actually the contribution of these forces to the chemical potential is given by the general equation (\ref{zeta}).  However, this equation cannot be used directly, because it gives the chemical potential $\zeta$ in terms of the density $\rho$, while (\ref{sat}) requires $\zeta$ as a function of the pressure $P$.
 
According to the conditions of mechanical equilibrium, the normal component $\sigma_{zz}$ of the stress tensor must be continuous at the surface of the film 
$-P_0(\rho,T)+\sigma^{(elm)}_{zz}=-P$. 
Then $\rho(P_0)=\rho(P+\sigma^{(elm)}_{zz}) \approx \rho(P)+(\partial \rho)/\partial P) \sigma^{(elm)}_{zz}$ and
$\zeta_0(\rho) \approx \zeta_0(P)+ (\partial \zeta/\partial \rho)(\partial \rho/\partial P) \sigma^{(elm)}_{zz}=
\zeta_0(P)+  \sigma^{(elm)}_{zz}/\rho$, where we took into account that $(\partial \zeta/\partial \rho)=1/\rho$. Thus we have
\begin{eqnarray}
\zeta(P,d)&=&\zeta_0(P)+\sigma^{(elm)}_{zz}/\rho+\frac{\hbar T}{4\pi} {\sum_{s=0}^{\infty}}~^{\prime} {\cal D}^{E}_{ll}(\mathbf{r},\mathbf{r};\xi_{s}) 
\frac{\partial \varepsilon(i\xi_{s};\mathbf{r})}{\partial \rho} \nonumber\\
&=&\zeta_0(P)+\sigma_{zz}'/\rho = \zeta_0(P)+F(d)/\rho\; ,
\label{zetaP}\end{eqnarray}
where $F(d)$ is the force, which in the London regime is given by (\ref{plates}) with 
$\varepsilon_2=1, l \rightarrow
 d$. (As far as electromagnetic properties of the vapour are concerned, we can treat it as a vacuum.) One can now rewrite  (\ref{sat})
in the form
\begin{equation}
F(d)=-\frac{T}{m}\ln\frac{P}{P_{sat}}
 \; .
\label{sat1}\end{equation}
If the film is placed on a solid wall situated vertically in the gravitational field, the dependence of the film thickness on the altitude is given by the equation
\begin{equation}
F(d)=\rho g x
 \; ,
\label{g}\end{equation}
where $x$ is the height.

In conclusion, let us consider a ``free'' film in vacuum. Then the chemical potential can be written as
\begin{equation}
\zeta(P,T,d)=\zeta_0(P,T)+F(d)/\rho
 \; 
\label{free}\end{equation}
where $F$ can be obtained from (\ref{plates}) with $\varepsilon_1=\varepsilon_2=1, l \to d$:
\begin{equation}
F=\frac{\hbar}{16\pi d^3}\int_0^{\infty}\int_0^{\infty}x^2\left 
[\frac{(1+\varepsilon)^2}{(1-\varepsilon)^2}e^x-1 \right ]
dxd\xi\; .
\label{freeplates}
\end{equation}
Note that this is just the quantity which can be calculated by integration of the force in the three-boundary geometry of Fig. 1. However, these calculations have never
been performed, and the correctness of the corresponding considerations has not been proved.

I thank R.~Scott for critical reading of this paper and useful suggestions.


\begin{thebibliography}{99}
\bibitem{footnote1} I use the generic term ``van der Waals forces'' for long-range forces between neutral objects in any conditions. Thus I do not distinguish between the London, Casimir, Casimir-Polder and Lifshitz  forces.
\end{thebibliography}
\end{document}